\documentclass[fleqn,twoside]{article}
\usepackage{espcrc2}

\usepackage{graphicx}
\usepackage[figuresright]{rotating}

\newcommand{\AmS}{{\protect\the\textfont2
  A\kern-.1667em\lower.5ex\hbox{M}\kern-.125emS}}

\title{Three Flavor QCD at High Temperatures}

\author{The MILC Collaboration: C.~Bernard
\address{Department of Physics, Washington University, St.~Louis, MO 63130, USA},
T.~Burch
\address{Universit\"at Regensburg, Theoretische Physik, 93040 Regensburg, Germany},
C.~DeTar
\address{Physics Department, University of Utah, Salt Lake City, UT 84112, USA},
Steven Gottlieb
\address{Department of Physics, Indiana University, Bloomington, IN 47405, USA},
E.B.~Gregory
\address{Department of Physics, University of Arizona, Tucson, AZ 85721, USA},
U.M.~Heller
\address{American Physical Society, One Research Road, Ridge, NY 11961--9000, USA},
J.~Osborn$\,\null^{\rm c}$,
R.L.~Sugar
\address{Department of Physics, University of California, Santa Barbara, CA 93106, USA}\thanks{Presented by R.L.~Sugar},
D.~Toussaint$\,\null^{\rm e}$
}
       
\begin{document}

\begin{abstract}
We have continued our study of the phase diagram of high temperature QCD with three 
flavors of improved staggered quarks. We are performing simulations with three degenerate 
quarks with masses less than or equal to the strange quark mass $m_s$ and with 
degenerate up and down quarks with masses $m_{u,d}$  less than the strange
quark mass.  For the quark masses studied to date, we find a crossover that strengthens 
as $m_{u,d}$ decreases, rather than a {\it bona fide} phase transition. We present 
new results for the crossover temperature extrapolated to the physical value
of $m_{u,d}$, and for quark number 
susceptibilities.

\vspace{1pc}
\end{abstract}

\maketitle

\section{THE PHASE DIAGRAM}

The MILC Collaboration is studying high temperature QCD with three flavors
of improved staggered quarks~\cite{EARLY} using 
the Asqtad quark action~\cite{ASQTAD}.
Simulations are being carried out with lattice spacings $1/4T$, $1/6T$
and $1/8T$.  We are considering two cases:
1) all three quarks have the same mass $m_q$; and 2) the two lightest
quarks have equal mass $m_{u,d}$, while the mass of the third
quark is fixed at approximately that of the strange quark
$m_s$. We refer to these cases as $N_f=3$ and $N_f=2+1$, respectively.
We have carried out studies with $0.2\, m_s\leq m_q\leq m_s$
for the $N_f=3$ case, and with $0.1\, m_s \leq m_{u,d} \leq m_s$ for
$N_f=2+1$.  At the masses we have studied to date, we find rapid crossovers,
which sharpen as the quark mass is reduced, rather than a {\it bona fide}
phase transition. During the past year we have extended and clarified our
calculations of the quark number susceptibilities, which
provide an excellent signal for the crossover, and which are directly
related to event by event fluctuations in heavy ion collisions.
Comparison of susceptibilities at different lattice spacings show that
our results are close to the continuum values.  We have also sharpened
our estimates of the crossover temperature for both the $N_f=3$ and
$N_f=2+1$ cases.

\begin{figure}
\centerline{\includegraphics[width=2.7in]{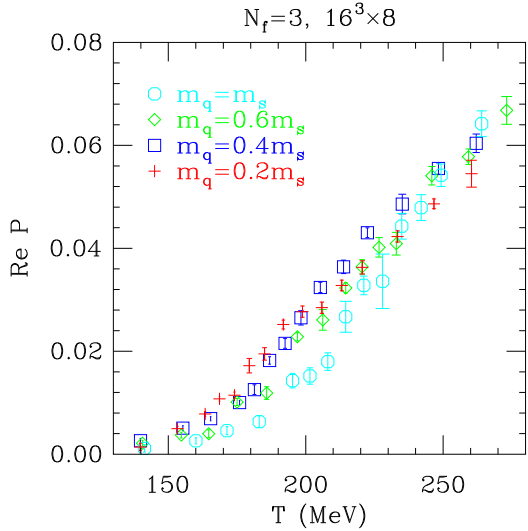}}
\vspace{-7mm}
\caption{The real part of the Polyakov loop  on $16^3\times 8$
lattices for $N_f=3$. 
\label{rp_nf3_nt8} }
\vspace{-4mm}
\end{figure}

For both the $N_f=3$ and $2+1$ cases we have carried out thermodynamics
studies on lattices with four, six and eight times slices, and aspect
ratio $N_s/N_t=2$. Here $N_s$ and $N_t$ are the spatial and temporal dimensions 
of the lattice in units of the lattice spacing. At the lightest quark masses,
we also performed some simulations with aspect ratio three, and obtained results
that are indistinguishable from those with aspect ratio two. For both cases
the standard thermodynamic quantities show a crossover from confined behavior at 
low temperature to deconfined behavior at high temperature, as is illustrated in 
Fig.~\ref{rp_nf3_nt8} where we plot the real part of the Polyakov loop
for $N_f=3$ on $16^3\times 8$ lattices. There is a slight 
trend for the temperature dependence of the Polyakov loop to be steeper for larger quark
masses. This is to be expected, since at sufficiently large quark masses,
it is a {\it bona fide} order parameter. The $\bar\psi\psi$ susceptibility 
$\chi_{\rm tot}$ 
provides a clear signal for the crossover. It is given by

\begin{equation}
\chi_{\rm tot} = \frac{\partial}{\partial m} \langle\bar \psi \psi\rangle.
\end{equation}

\begin{figure}
\centerline{\includegraphics[width=2.7in]{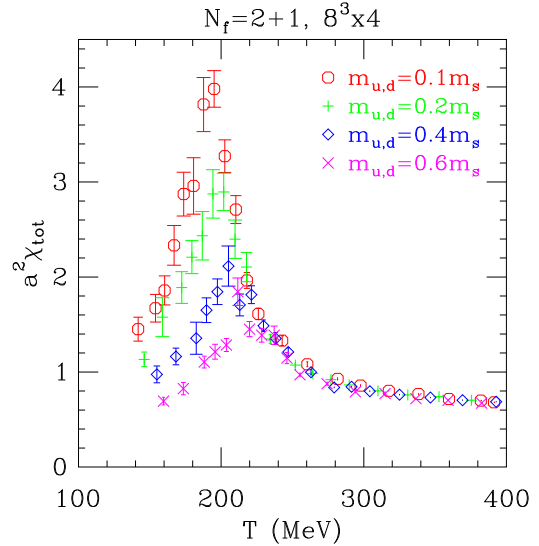}}
\vspace{-7mm}
\caption{The $\bar\psi\psi$ susceptibility as a function of temperature
on $8^3\times 4$ lattices for $N_f=2+1$.
\label{chi_tot_nf21_nt4}}
\vspace{-4mm}
\end{figure}

\noindent
We plot this quantity as a function of temperature on $8^3\times 4$ lattices in 
Fig.~\ref{chi_tot_nf21_nt4}. Note the increase in the 
height of the peak as the quark mass is decreased.
Our work and that of the Bielefeld group~\cite{P4MASS} strongly
suggests that in the real world, $N_f=2+1$, there is no phase
transition at the physical quark masses. With this assumption, we have estimated
the critical temperature for $N_f=2+1$ at the physical value of $m_{u,d}$ 
through an extrapolation of the form

\begin{equation}
r_1\,T_c=c_0+c_1\,(m_\pi/m_\rho)^d + c_2\, (aT_c)^2,
\label{eq:r1Tc}
\end{equation}

\noindent
where we evaluated $T_c$ for each value of $N_t$ and $m_\pi/m_\rho$ for which
we have made measurements from the peak in the $\bar\psi\psi$ susceptibility.
For a second order phase transition in the O(4) universality class at
$m_{u,d}=0$, $d=2/\beta\delta\approx 1.08$. 
We find that  $T_c=169(12)(4)$~MeV with a $\chi^2$ of 2.1 for 11 degrees
of freedom. The first error is the fit error, the second from the uncertainty
in $r_1$, taken as 0.317(7)~fm~\cite{R1}.  
To test the sensitivity of $T_c$ to $d$, we have also performed
a fit with $d=2$, which yields $T_c=174(11)(4)$~MeV with a $\chi^2$ of 1.5 for 11 
degrees of freedom. So, the goodness of the fit does not allow us to prefer either
of them.

\section{QUARK NUMBER SUSCEPTIBILITIES}

In order to study the quark number
susceptibilities~\cite{SUSC,GG}, we introduce
chemical potentials $\mu_\alpha$ coupled to a set of mutually
commuting conserved charges $Q_\alpha$. 
The quark number susceptibilities are related to event-by-event
fluctuations in heavy ion collisions~\cite{EbyE} by the
fluctuation-dissipation theorem

\begin{equation}
\chi_{\alpha,\beta}(T) =
\left\langle(Q_\alpha-\langle Q_\alpha\rangle) (Q_\beta-\langle Q_\beta\rangle) \right\rangle.
\label{eq:chi}
\end{equation}

\noindent
We work at $\mu_\alpha=0$, so the brackets, $\langle\rangle$, in
Eq.~(\ref{eq:chi}) indicate averages weighted by the standard, real
Euclidean action for QCD, and $\langle Q_\alpha\rangle=0$.

\begin{figure}
\centerline{\includegraphics[width=2.7in]{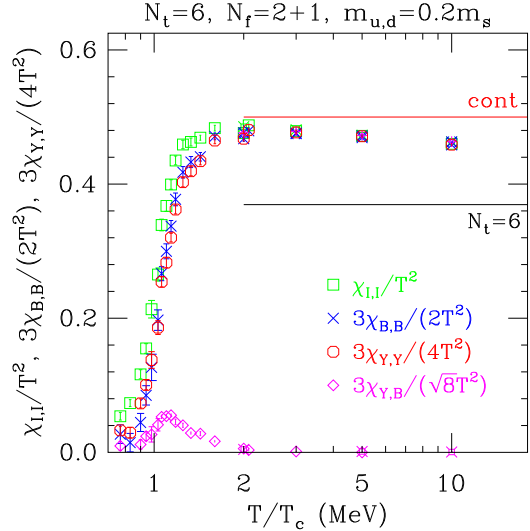}}
\vspace{-7mm}
\caption{The non-vanishing elements of the susceptibility matrix
for two light quarks of mass $0.2\, m_s$ and one heavy quark of mass
$m_s$ on $12^3\times 6$ lattices.
\label{qno_combo_new2} }
\vspace{-4mm}
\end{figure}

Rather than choosing the three independent charges to be the 
number operators for up, down and strange quarks, it appears more physical to take 
them to be the $z$-component of isospin, $Q_I$, the hypercharge, $Q_Y$, and the 
baryon number, $Q_B$. 
The rows and columns of the susceptibility matrix, $\chi_{\alpha,\beta}(T)$ are
then labeled by $I$, $Y$ and $B$. In the $N_f=3$ case,
where $m_u=m_d=m_s$, $\chi$ is a diagonal matrix, and there are no correlations
among fluctuations in $Q_I$, $Q_Y$ and $Q_B$, while for the $N_f=2+1$ case, 
where $m_u=m_d$,  the only correlations are between fluctuations in hypercharge 
and baryon number. For temperatures below the phase transition or 
crossover, the lightest particle that can be excited by a chemical potential 
coupled to $Q_I$ is the pion, while for chemical potentials coupled to
$Q_Y$ and $Q_B$ it is the kaon and the nucleon, respectively.
Above the transition temperature each of the chemical potentials
can excite quark states that are much lighter than hadrons, so
we expect the diagonal elements of $\chi$ to increase sharply in
the vicinity of the transition, and they do.
This is illustrated in Fig.~\ref{qno_combo_new2} where
we plot the diagonal elements of the susceptibility matrix,
$\chi_{I,I}$, $\chi_{Y,Y}$ and $\chi_{B,B}$, as a function of
temperature for two light quarks with mass $0.2\, m_s$ and one
heavy quark with mass $m_s$ on $12^3\times 6$ lattices. $\chi_{Y,Y}$
and $\chi_{B,B}$ have been multiplied by factors of 3/4 and 3/2
respectively, so that the quantities plotted approach the same high
temperature limit as $\chi_{I,I}$. Also shown is $\chi_{Y,B}$,
the only non-zero off-diagonal matrix element of $\chi$ for
$m_u=m_d$. It measures correlations between fluctuations in the
hypercharge and baryon number. The coefficient of $\chi_{Y,B}$ in this
figure is the geometric mean of those for $\chi_{Y,Y}$ and $\chi_{B,B}$.
The good agreement between these results, and similar ones on lattices
with eight time slices, indicate that they are close to their physical values.

This work is supported by the US National Science Foundation and
Department of Energy and used computer resources at Florida State
University (SP), Indiana University, NCSA, NERSC, NPACI (Michigan), 
FNAL, and the University of Utah (CHPC).

\end{document}